\documentclass[fleqn,usenatbib]{mnras}

\usepackage{showyourwork}

\usepackage{mathptmx}

\usepackage[T1]{fontenc}

\DeclareRobustCommand{\VAN}[3]{#2}
\let\VANthebibliography\thebibliography
\def\thebibliography{\DeclareRobustCommand{\VAN}[3]{##3}\VANthebibliography}

\usepackage{graphicx}	
\usepackage{amsmath}	
\usepackage{siunitx}

\title[The dispersal of compact protoplanetary discs]{The dispersal of compact protoplanetary discs}

\author[G. Picogna et al.]{
Giovanni Picogna$^{1}$\thanks{E-mail: picogna@usm.lmu.de}
and Barbara Ercolano$^{1,2,3}$
\\
$^{1}$Universit\"{a}ts-Sternwarte, Ludwig-Maximilians-Universit\"{a}t M\"{u}nchen, Scheinerstr. 1, D-81679 M\"{u}nchen, Germany\\
$^{2}$Excellence Cluster Origins, Boltzmannstrasse 2, D-85748 Garching bei M\"{u}nchen, Germany\\
$^{3}$Max-Planck-Institut f\"{u}r Extraterrestrische Physik, Giessenbachstr. 1, 85748 Garching, Germany
}

\date{Accepted XXX. Received YYY; in original form ZZZ}

\pubyear{2025}

\begin{document}

\pagerange{\pageref{firstpage}--\pageref{lastpage}}
\maketitle

\begin{abstract}
\label{firstpage}
Compact protoplanetary discs are becoming increasingly prominent in observations. Their dispersal pathways may differ substantially from those of extended discs. We aim to quantify the role of the disc outer radius in internal photoevaporation, provide a simple scaling relation for compact discs, and test whether the resulting evolutionary tracks reproduce the observed inside-out clearing of young stellar populations. We performed radiation-hydrodynamic simulations of X-ray-driven photoevaporation for discs with different outer radii, and derived the dependence of the total mass-loss rate on the cut-off radius. We find that the surface mass-loss profiles are nearly independent of disc size, but their integrated wind rates are reduced according to the cumulative mass-loss rate distribution. We incorporated this scaling into disc population synthesis models. When the internal photoevaporation is applied only up to the cut-off radius compact discs evolve via inside-out clearing consistent with observational diagnostics, while when the cut-off radius is not considered, the disc spreading is hindered and the disc dispersal proceeds from the outside-in. The introduction of mild external photoevaporation present in nearby star forming regions cannot prevent the disc spreading when the cut-off radius prescription is included, but it can much better explain the evolution of disc radii as a function of time. Disc dispersal prescriptions must include the dependence on disc cut-off radius to capture the evolution of compact discs. The proposed scaling provides a simple, physically motivated correction that better predicts the growing observational evidence for compact discs and inside-out dispersal.
\end{abstract}




\section{Introduction}\label{sec:introduction}

 Protoplanetary discs are the birthplaces of planets, and their lifetimes and dispersal pathways set essential constraints on planet formation models. Most discs dissipate within a few Myr, with only a minority surviving to \SI{\sim 10}{Myr} \citep{2001ApJ...553L.153H, 2010A&A...510A..72F, 2020ApJ...890..106S, 2022ApJ...939L..10P}. Observations also show that dispersal proceeds preferentially from the inside-out: colour-colour evolution studies of young stellar populations \citep{2011MNRAS.410..671E, 2013MNRAS.428.3327K, 2015MNRAS.452.3689E} reveal clear signatures of inner clearing, inconsistent with outside-in erosion. Any successful theoretical model must therefore reproduce this inside-out behaviour.

 Mounting evidence points to magnetically driven winds and photoevaporation (EUV, X-ray, FUV, both internal and external) as the dominant agents of angular momentum and mass loss \citep{2017RSOS....470114E, 2023ASPC..534..465L, 2023ASPC..534..567P}. X-ray photoevaporation models in particular have been refined in recent years to yield dispersal prescriptions consistent with several independent observational diagnostics \citep{2016MNRAS.460.3472E, 2020MNRAS.496..223W, 2021A&A...648A.121F, 2023ApJ...955L..11R, 2023MNRAS.526L.105E}.

 Most existing models, however, assume discs with outer radii of order hundreds of au - well beyond the gravitational radius where the gas sound speed becomes comparable with the Keplerian speed, creating a natural "sweet spot" for photoevaporation \citep{1994ApJ...428..654H}
 \begin{equation}
    R_g = \frac{GM_\star}{c_s^2}\,.
 \end{equation} 
 One can get a better estimate of the gravitational radius by considering the Bernoulli parameter (see eq.~\ref{eq:Bernoulli}) leading to \citep[for a full derivation see][]{2003PASA...20..337L}:
 \begin{equation}
   R_g = \frac{\gamma-1}{2\gamma} \frac{GM_\star \mu m_H}{k_B T}\,.
 \end{equation}
 If the gas is fully ionized, one can assume a temperature of $T = 10^4$ K for a pure EUV wind \citep{2012MNRAS.422.1880O}, with adiabatic index $\gamma=5/3$, and mean molecular weigth $\mu = 0.61$ leading to a gravitational radius of $1.31$ au for a solar mass star. However, recent studies have pointed towards more cool molecular winds \citep[see e.g.][]{2024A&A...690A.296S}, where the temperature at the base of the flow is non-uniform and lower than $10^4$ K, leading to a larger gravitational radius. For example, assuming a temperature of $T = 5000$ K, with adiabatic index $\gamma=7/5$, and mean molecular weigth $\mu = 2.35$, one gets $R_g = 7.22$ au. In order not to keep this as an additional free parameter, we adopt a value of $R_g = 5$ au as a representative scale for the gravitational radius of a solar mass star in the rest of this work.
 
 Yet compact discs, with outer radii comparable to or smaller than this scale, are increasingly recognised as common.
 ALMA surveys show that a substantial fraction of discs are indeed very compact \citep[e.g.][]{Trapman_2025}, with radii much smaller than typically assumed even at an evolved stage. 
 Such compact configurations may result from internal evolution (mass and angular momentum loss via winds, or a close companion) or from external influences such as stellar fly-bys or external photoevaporation. 
 In particular, external photoevaporation (even at a moderate level) can shrink considerably the discs on a short timescale ($< 1$ Myr), until an equilibrium is reached between the viscous spreading and the external photoevaporation \citep{2020MNRAS.492.1279S, 2026arXiv260222050P}. After that the disc radius does not evolve significantly for the rest of the disc life-time.

 In this paper, we investigate the dispersal of compact discs by explicitly accounting for their finite outer radii in photoevaporation models.
 Using radiation-hydrodynamic simulations, we derive the dependence of the total wind mass-loss rate on the cut-off radius and implement this scaling into one-dimensional viscous evolution and population synthesis models of compact discs.
 Our aim is to provide a simple prescription valid in the compact-disc regime and to test whether the resulting evolutionary pathways are consistent with observed disc fractions, accretion rates, and the observationally inferred inside-out dispersal of young stellar populations.

\section{Methods}\label{sec:methods}

We performed a series of radiation hydrodynamic simulations following the approach described in \citet{2019MNRAS.487..691P}, which we briefly summarise here for completeness.

\subsection{Radiative Transfer}

We focused on a \SI{1}{M_\odot} star, where the stellar properties were taken from \citet{2000A&A...358..593S}, assuming an age of 1 Myr and metallicity $Z=0.02$ without convective overshooting.
The initial disc structure was based on the gas densities and dust temperatures from the hydrostatic disc models of the D'Alessio Irradiated Accretion disc (\textsc{diad}) radiative transfer framework \citep{1998ApJ...500..411D, 1999ApJ...527..893D, 2001ApJ...553..321D, 2005ApJ...621..461D, 2006ApJ...638..314D}. These models provide the best fits to the median spectral energy distribution (SED) observed in Taurus.
To compute the thermal structure of the upper layers of the disc, we used the gas photoionisation and dust radiative transfer code \textsc{mocassin} \citep{2003MNRAS.340.1136E,2005MNRAS.362.1038E,2008ApJS..175..534E}.
This code self-consistently solves for heating and cooling processes under thermal equilibrium, allowing us to derive temperature prescriptions up to the maximum penetration depth of X-rays ($\sim 10^{22}$ pp/cm$^2$) for discs irradiated by X-ray and EUV stellar spectra.
The resulting temperature prescription depends on the spectral hardness as detailed in \citet{2021MNRAS.508.1675E}.
It relates the local gas temperature to both the column density towards the central star (from \SI{5e20}{pp.cm^{-2}} to \SI{2e22}{pp.cm^{-2}}), and the local ionisation parameter \citep{1969ApJ...156..943T} $\xi = L_X/(n r^2)$, where $L_X$ is the stellar X-ray luminosity, $n$ the local number density, and $r$ the spherical radius.

\subsection{Hydrodynamical model}

Having the temperature prescription at thermal equilibrium, we then performed hydrodynamical simulations using a modified version of the \textsc{pluto} code \citep{2007ApJS..170..228M}, as described in \citet{2019MNRAS.487..691P}.
In this setup, the \textsc{mocassin}-derived prescription was applied at column densities below the maximum X-ray penetration depth \citep[\SI{\sim 1e22}{cm^{-2}}, see e.g. Fig.~3 of][]{2009ApJ...699.1639E}, while at larger column densities we assumed perfect thermal coupling between gas and dust, adopting the \textsc{diad} temperature prescription.
Each simulation was evolved until the disc structure and wind streamlines reached steady state.
We explored $4$ different initial cut-off radii for the circumstellar disc: $10$, $50$, $100$, $200$ au, where we exponentially cut off the gas surface densities from the unperturbed disc.
We adopted a spherical coordinate system centred on the star.
The computational grid is logarithmically spaced in the radial direction with $1024$ cells, ensuring enhanced resolution in the innermost disc regions where photoevaporation is most effective, while still extending to large radii ($R_\mathrm{out} = 300\,\mathrm{au}$ for the more compact discs and $R_\mathrm{out} = 1,000\,\mathrm{au}$ for the remaing) without excessive computational cost. In the polar direction, the grid is linearly spaced with $256$ cells between \SI{0.01}{rad} and $\pi/2$.
Outflow boundary conditions were applied both at the inner and outer radial boundaries, while reflecting boundary conditions were used along the disc midplane and special axisymmetric boundary condition at the polar axis (for more information see \citet{2007ApJS..170..228M}).
The systems were evolved for $\sim 500$ orbits at \SI{10}{au}, adopting an $\alpha = 0.001$.
A quasi-steady state is typically reached after a few hundred orbital periods, at which point the cumulative mass-loss rate and the gas streamlines in the wind stabilise.

\subsection{Disc population synthesis}

In order to model the long-term evolution of compact protoplanetary discs, we adopted a 1D viscous evolution code \citep{2015MNRAS.450.3008E} where the 1D surface density is evolved according to
\begin{equation}\label{eq:visc}
\frac{\partial\Sigma}{\partial t} = \frac{1}{R} \frac{\partial}{\partial R} \left[ 3R^{1/2} \frac{\partial}{\partial R} \left( \nu\Sigma R^{1/2} \right)\right] - \dot{\Sigma}_{\mathrm{w_{in}}}(R, t) - \dot{\Sigma}_{\mathrm{w_{out}}}(R, t)\,,
\end{equation}
where the first term on the right-hand side describes the disc viscous evolution \citep{1974MNRAS.168..603L}, the second term the mass loss due to photoevaporation derived from our hydrodynamical models \citep{2021MNRAS.508.3611P}, and the last term the mass loss due to external photoevaporation \citep{2023MNRAS.526.4315H}.

\subsubsection{Internal Photoevaporation prescription}
The mass-loss rate due to internal photoevaporation is given by \citep{2021MNRAS.508.3611P}
\begin{eqnarray}
  \label{eq:surf1}
  \dot{\Sigma}_{\mathrm{w_{in}}} &= \ln{(10)} \bigg(\frac{6\, a\, \ln{(R)}^5}{R\, \ln{(10)}^6} +
  \frac{5\, b\, \ln{(R)}^4}{R\, \ln{(10)}^5} +
  \frac{4\, c\, \ln{(R)}^3}{R\, \ln{(10)}^4} + \\ \nonumber
  &\frac{3\, d\, \ln{(R)}^2}{R\, \ln{(10)}^3} +
  \frac{2\, e\, \ln{(R)}}{R\, \ln{(10)}^2} + \\ \nonumber
  &\frac{f}{R\, \ln{(10)}}\bigg)
  \frac{\dot{\mathrm{M}}_w(R)}{2\pi\, R} \ [\SI{}{M_\odot \textrm{au}^{-2} yr^{-1}}]\,,
\end{eqnarray}
where
\begin{equation}\label{eq:mdot}
    \dot{M}_{w} = A(R) B(L_X) \dot{M}_w(M_\star) \ [\SI{}{M_\odot yr^{-1}}]\,,
\end{equation}
is the total mass-loss rate due to photoevaporation.

On the right hand side, the first term represents the normalized cumulative mass-loss rate as a function of radius \citep{2021MNRAS.508.3611P}
\begin{equation}
\label{eq:surf2}
  A(R) = 10^{a\log{R}^6 + b\log{R}^5 + c\log{R}^4 + d\log{R}^3 + e\log{R}^2 + f\log{R} + g}
\end{equation}
where the parameters $a$, $b$, $c$, $d$, $e$, $f$, $g$ are stellar-mass dependant and given in \citet[][Table 2]{2021MNRAS.508.3611P}.
The second term is the normalized mass-loss rate dependence on the X-ray luminosity \citep{2021MNRAS.508.1675E}
\begin{equation}\label{eq:mdotlx}
    \begin{split}
    B(L_X) &= \frac{\dot{M}_w(L_{X,\mathrm{soft}} / \mathrm{erg} \,\mathrm{s}^{-1})}{\dot{M}_w(L_{X,\mathrm{soft, mean}}/ \mathrm{erg} \,\mathrm{s}^{-1})} \\
    &= 10^{a_L \left[\exp{\left(\frac{(\ln(\log(L_{X,\mathrm{soft}})-b_L)^2}{c_L}\right)} -\exp{\left(\frac{(\ln(\log(L_{X,\mathrm{soft, mean}})-b_L)^2}{c_L}\right)}\right]} \,,
    \end{split}
\end{equation}
with $a_L = -1.947\cdot 10^{17}$, $b_L = -1.572\cdot 10^{-4}$, $c_L = -0.2866$, the soft component of the X-ray luminosity is
\begin{equation}\label{eq:lxsoft}
    L_{X,\mathrm{soft}} = 10^{0.95 \log{(L_X / \mathrm{erg} \,\mathrm{s}^{-1})}+1.19} \ [\SI{}{erg \, s^{-1}}]\,,
\end{equation}
and $L_{X,\mathrm{soft, mean}}$ is the soft component for a star with total X-ray luminosity given by the observational relation between stellar mass and mean X-ray luminosity \citep{2007A&A...468..353G}
\begin{equation} \label{eq:LxMstar}
    L_{X, \mathrm{mean}} = 10^{1.54 \log(M_\star / M_\odot) + 30.31} \ [\SI{}{erg \, s^{-1}}]\,.
\end{equation}

The third term is the mass-loss rate as a function of stellar mass (for a mean X-ray luminosity, \citet{2021MNRAS.508.3611P})
\begin{equation}\label{eq:mdotmstar}
    \dot{M}_w(M_\star) = 3.93\times10^{-8} \left(\frac{M_\star}{M_\odot}\right) \left[\frac{M_\odot}{\mathrm{yr}}\right] \ [\SI{}{M_\odot yr^{-1}}]\,.
\end{equation}

The absolute values of photoevaporative mass-loss rates remain uncertain, as they depend sensitively on several poorly constrained parameters - including the gas metallicity \citep[see e.g.][who show that modest C depletions within the discs can lead to order magnitude increase in mass loss rates]{2019MNRAS.490.5596W}, the hardness and luminosity of the stellar X-ray spectrum \citep{2021MNRAS.508.1675E}, the accretion luminosity and the corresponding heating-cooling balance in the flow.
However, \citet{2024A&A...690A.296S} showed that the lack of O-H collisional cooling, missing in the previous models \citep{2012MNRAS.422.1880O, 2019MNRAS.487..691P}, were reducing by an order of magnitude the mass-loss rates, for the same parameter choice.
Our main goal in this work is not to reassess the absolute normalisation of the rates, but rather to explore their dependence on disc outer radius and its impact on the evolutionary behaviour of compact discs.
Nevertheless, to facilitate comparison with the most recent results in the literature \citep[e.g.][]{2024A&A...690A.296S}, we have repeated all our calculations using a mass-loss prescription scaled by a factor of 10 with respect to \citet{2019MNRAS.487..691P}.
This choice follows the current convention adopted in population synthesis and evolutionary models to better fit observational data. We emphasize, however, that such a uniform scaling may not fully capture the effects of chemistry and metallicity, which also modify the radial profile of the mass-loss rate \citep{2019MNRAS.490.5596W,2024A&A...690A.296S}.
Throughout the paper, we consider the discs to be dispersed when the maximum surface density drops below $\Sigma_\mathrm{min} = 10^{-2}$ g/cm$^2$. This condition should mock a sensitivity limit for the detection of gas in discs.

\subsubsection{External Photoevaporation prescription}

We implemented external photoevaporation by far-ultraviolet (FUV) radiation from nearby OB stars using version 2 of the \textsc{fried} grid \citep{2018MNRAS.481..452H, 2023MNRAS.526.4315H}.
This grid tabulates the total FUV-driven disc mass-loss rate $\dot{M}_\mathrm{FUV}$ as a function of stellar mass $M_\star$, disc outer radius $R_\mathrm{disc}$, surface density at 1~au $\Sigma_{1\,\mathrm{au}}$, and FUV field strength $G_0$.
Because photoelectric emission from polyciclic aromatic hydrocarbons (PAHs) constitutes the primary heating mechanism within the photodissocation region (PDR), the assumed PAH abundance is a critical parameter for these rates \citep{2016MNRAS.457.3593F}.
Curiosly, PAHs are commonly observed in discs around massive Herbig stars but remain elusive in T-Tauri discs \citep[e.g.][]{2017ApJ...835..291S, 2006A&A...459..545G, 2007A&A...466..229V}.
This lack of detection likely stems from three possible factors: (1) PAH destruction driven by strong X-ray emission from T-Tauri stars \citep{2010A&A...511A...6S}; (2) the loss of infrared signatures due to chemical processing into complex organics; or (3) insufficient sensitivity and wavelength coverage in current observational facilities.
Lacking direct observational constraints for our specific targets, we adopt an ISM-like PAH-to-dust abundance ratio of unity as our fiducial baseline, reflecting an ISM-like PAH population that scales with ISM-like dust.

First we select the closest stellar mass grid point (which are $0.1, 0.3, 0.6, 1.0, 1.5$ and $3 M_\odot$) to our target star, since it wouldn't be physically consistent to interpolate between different stellar masses as the disc properties in the radiative transfer calculations were also changed accordingly \citep[for more details, see][]{2023MNRAS.526.4315H}.
Then, at each timestep, we compute the per-cell mass-loss rate following the prescription of \citet{2020MNRAS.492.1279S, 2024A&A...681A..84G}.
For every disc annulus at radius $R$ we interpolate the \textsc{fried} grid at the local parameters $(R,\,\Sigma(R),\,G_0)$ using a three-dimensional regular-grid interpolator in $(\log R,\,\log\Sigma_{1\,\mathrm{au}},\,\log G_0)$ space.
Here, $\Sigma(R)$ is the local surface density at radius $R$ assuming a power-law profile with index $-1$.
This approach allows us to continuously sample the FUV mass-loss rate as a function of radius, avoiding discrete binning onto the \textsc{fried} $G_0$ nodes.

The truncation radius, $R_\mathrm{trunc}$, marks the transition between the optically thick disc and the optically thin photoevaporative wind. We identify this radius as the outermost peak of the interpolated rate profile:
\begin{equation}
    R_\mathrm{trunc} = \underset{R}{\mathrm{argmax}}\;\dot{M}_\mathrm{FUV}(R).
\end{equation}
The total external mass-loss rate is then defined as $\dot{M}_\mathrm{ext} = \dot{M}_\mathrm{FUV}(R_\mathrm{trunc})$.

Following the numerical approach of \citet{2020MNRAS.492.1279S}, this total mass-loss rate is spatially partitioned across the individual outer grid cells. Specifically, mass is removed from all cells at $R \geq R_\mathrm{trunc}$ in proportion to the local surface density. The mass-loss rate for a given outer cell $i$ is weighted by its local mass fraction within the outer disc region:
\begin{equation}
  \dot{M}_{\mathrm{ext}, i} = -\dot{M}_\mathrm{ext} \frac{M_i}{M(R \geq R_\mathrm{trunc})},
\end{equation}
where $M_i$ is the mass in cell $i$, and $M(R \geq R_\mathrm{trunc})$ is the total disc mass beyond the truncation radius. This formulation ensures that the total integrated removal rate equals $\dot{M}_\mathrm{ext}$. Finally, for discs extending beyond the outer edge of the \textsc{fried} grid ($500$~au for version~2), material is stripped at this same total rate, allowing us to accurately handle large initial discs.

\subsubsection{Initial conditions}
The discs are initialized with a self-similar solution (for $\gamma = 1$):
\begin{equation}
    \Sigma = \frac{M_0}{2 \pi r_1 R} T^{-3/2} \exp{\left[-\frac{(R/r_1)}{T}\right]}\,,
\end{equation}
where $M_0$ is the initial disc mass, $T=t/t_s$ is the dimensionless time with $t_s$ the viscous timescale, and $r_1$ is the disc cut-off radius.

We performed a parameter space analysis by probing a large set of disc and stellar properties.

We assumed a stellar initial mass function following \citet{2001MNRAS.322..231K}, $\xi(m)\propto m^{-\alpha}$, where $\alpha = 1.3 \pm 0.5$ for $0.08 \leq m/M_\odot < 0.5$ and $\alpha = 2.3 \pm 0.3$ for $0.5 < m/M_\odot < 1$.

We derived an integrated initial multiplicity fraction of $41.3\%$ for this population, calculated as the weighted average of the observed fractions for M-dwarfs ($\sim 35 \%$) and solar-type stars \citep[$\sim 70 \%$, see e.g.][]{2013ARA&A..51..269D}. 
We further assumed that intermediate separation binaries ($3\leq a \leq 30$ au) are the most destructive to protoplanetary discs, as this regime corresponds to the typical locations of planet-forming reservoirs and leads to rapid tidal truncation \citep[see e.g.][]{2009ApJ...696L..84C, 2012ApJ...745...19K}.
By integrating the log-normal orbital period distributions appropriate for M-dwarfs \citep[peak at $a_b\simeq 20$ au][]{2019AJ....157..216W} and solar-type stars \citep[peak ay $a_b \simeq 50$ au][]{2010ApJS..190....1R}, we estimate that $29.9 \%$ of binaries in our sample fall within this destructive window.
This yields an effective initial disc fraction of $\sim 87.6 \pm 0.2 \%$, providing a physically motivated initial condition for the subsequent population synthesis.

For the X-ray luminosities we followed the same approach outlined in \citet{2023MNRAS.526L.105E}, deriving an X-ray luminosity function in three stellar mass bins from a subsample of the Chandra Orion Ultra-deep Project \citep[COUP, cf.][]{2005ApJS..160..319G} as shown in \citet{2019ApJ...883..117K}.

When included, the FUV field strength $G_0$ for each synthetic disc is drawn from the empirical distribution of disc-hosting stars in nearby star-forming regions ($d < 200$~pc; Upper~Sco, Taurus, Lupus, $\rho$~Oph, Cham~I/II, CrA) measured by \citet{2025A&A...695A..74A}, using a kernel density estimate in $\log G_0$. This allow us to compare the resulting population synthesis with the high quality sample of observed stellar and disc properties by \citet{2023ASPC..534..539M}.

For the disc properties, we derive the disc dust mass from observational constraint in young star forming regions \citep{2016ApJ...831..125P, 2017AJ....153..240A}:
\begin{equation}
    \log(M_\mathrm{dust} [M_\oplus]) = (1.2\pm0.2) + (1.8\pm0.4) \log(M_\star [M_\odot])
\end{equation}
with an associated uncertainty of $\delta = 0.9\pm0.1$.
We then consider a standard 100 conversion factor between the dust and the gas mass, and derive the cut-off radius by inverting the observed dependence of the cut-off radius as a function of the gas mass \citep{Trapman_2025}, $M_\mathrm{gas} \propto r_1^{1.7}$. We limit our initial sample to discs with cut-off radii between $1$ and $500$ au in order to be consistent with observational constraints. Finally, the disc viscosity is the less constrained parameter, thus we linearly sampled its values between $\alpha = 10^{-4}$ and $10^{-2 }$.

\begin{figure*}
    \centering
    \includegraphics[width=17cm]{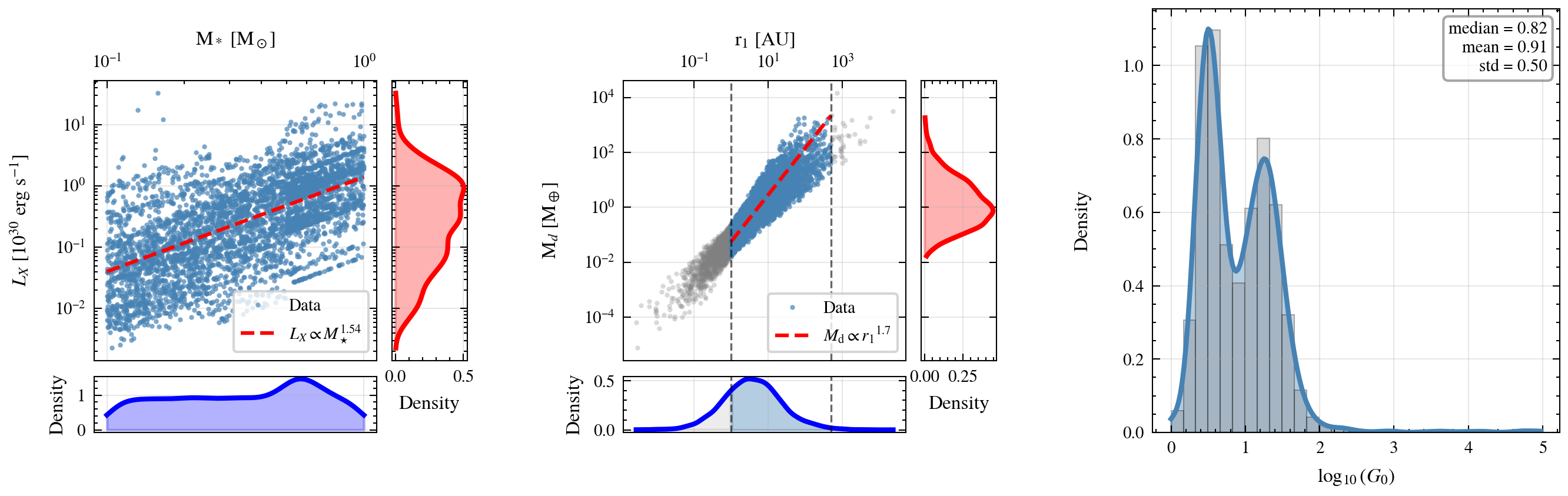}
    \caption{Parameter space sampled in the population synthesis. On the left panel, stellar mass as a function of X-ray luminosity sampled. In the middle panel, cut-off radius as a function of disc mass sampled. On the right panel, external FUV field strength $G_0$ sampled \citet{2025A&A...695A..74A}. The observationally derived fits from \citet{2007A&A...468..353G} and \citet{Trapman_2025} are overplotted for comparison.}
    \script{fig1.py}
    \label{fig:pop_synth_corner_plots}
\end{figure*}

To efficiently explore the 9-dimensional parameter space governing initial stellar and disc properties, we employed a Quasi-Monte Carlo (QMC) sampling strategy utilizing Sobol sequences \citep{sobol1967distribution}.
Unlike standard pseudo-random Monte Carlo sampling, which exhibits $\mathcal O(N-1/2)$ convergence and often results in clustering or gaps, Sobol sequences are low-discrepancy sequences that ensure a more uniform coverage of the hypercube with a convergence rate closer to $\mathcal O(N-1)$. The generated samples were mapped from the unit hypercube to their respective physical priors. For the dust mass scatter term, we applied an inverse cumulative distribution function to convert the uniform Sobol samples into a normal distribution $N(0,\delta_{Mdust})$. This approach ensures that our synthetic population of $4096$ systems provides a statistically robust representation of the parameter space.
The resulting parameter space sampled is shown in Fig.~\ref{fig:pop_synth_corner_plots}.

\section{Results}\label{sec:results}

\subsection{Hydrodynamical models}

\begin{figure*}
    \includegraphics{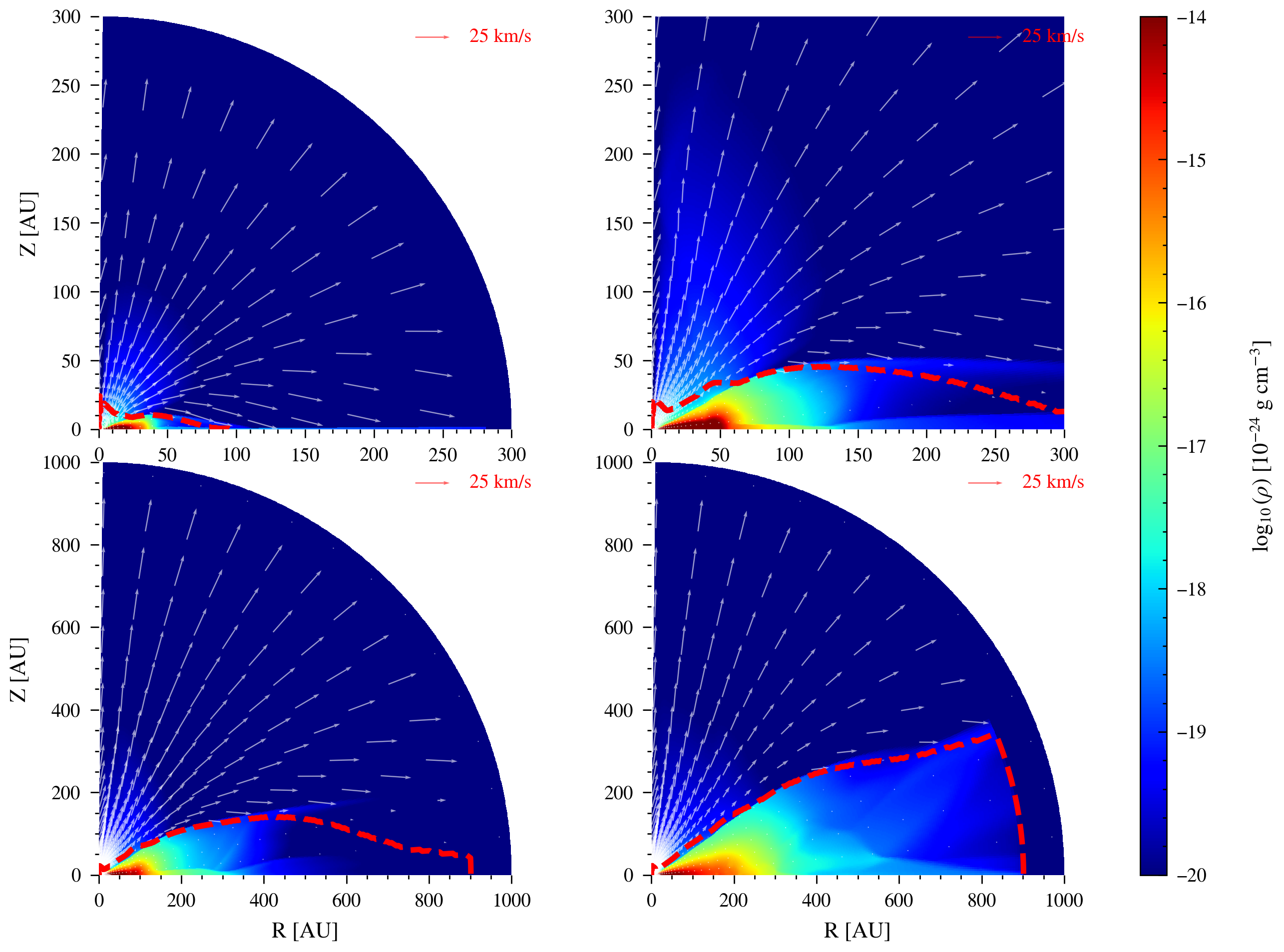}
    \caption{Density distribution and velocity field averaged over 10 orbits for the different cut-off radii (top row: 10 and 50 au, bottom row: 100 and 200 au). The dashed red line indicates the region where the Bernoulli parameter is zero, separating bound from unbound material.}
    \script{fig2.py}
    \label{fig:densityandvelocity}
\end{figure*}

\begin{figure}
	\includegraphics{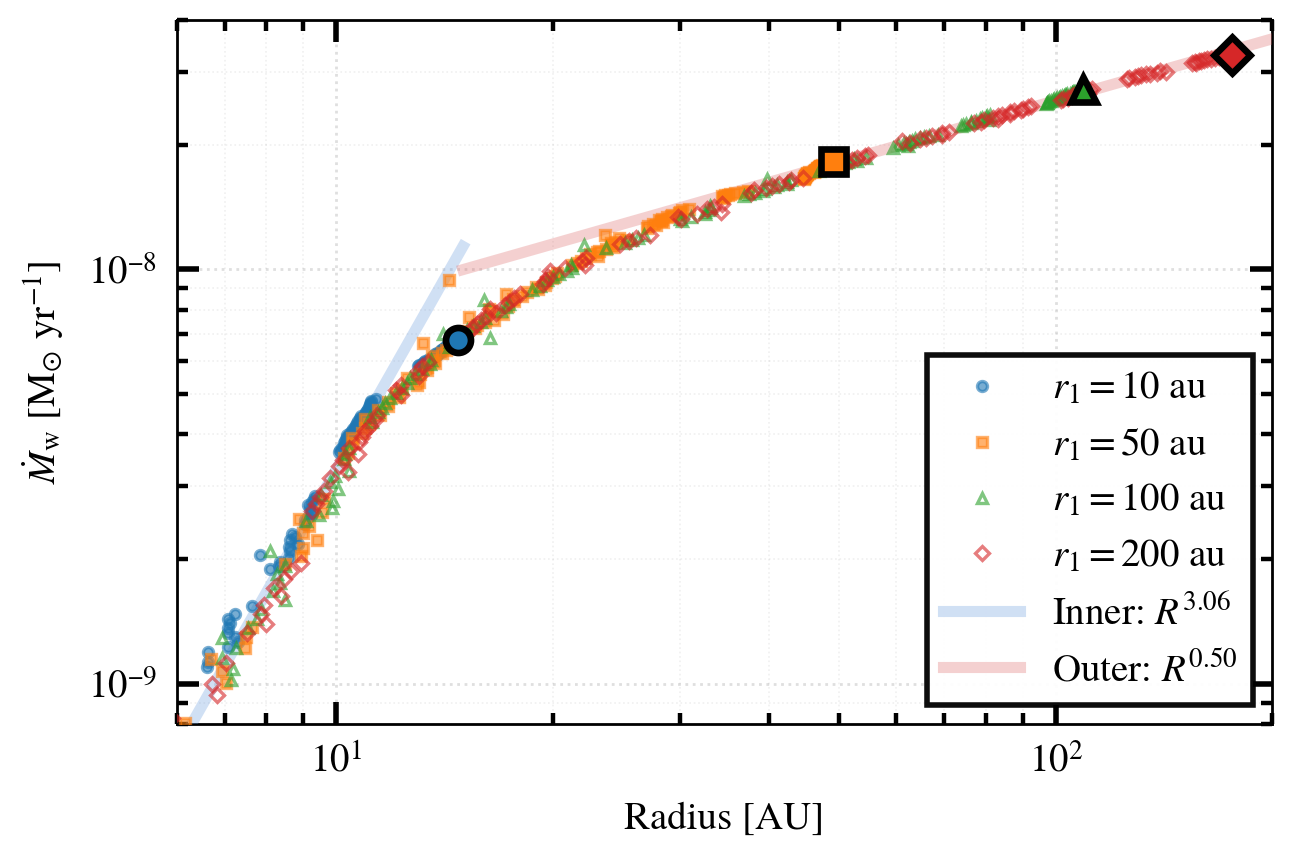}
	\caption{Cumulative mass loss rate as a function of radius for different cut-off radii.}
        \script{fig3.py}
	\label{fig:mass_loss_rate_cumulative}
\end{figure}

We ran four hydrodynamical simulations with different cut-off radii, as described in Section~\ref{sec:methods}.
The resulting density and velocity structure of the wind are shown in Fig.~\ref{fig:densityandvelocity}.
The overall wind morphology is remarkably consistent across the different disc sizes, as the outer disc extent does not affect the inner disc regions where the wind is launched.
The cut-off radius does not evolve significantly over the simulation time, as the timescales considered are short compared to the viscous timescale at the outer edge.
We computed the region where the disc is gravitationally bound by computing the Bernoulli parameter \citep{2003PASA...20..337L}:
\begin{equation} \label{eq:Bernoulli}
    \mathcal{B} = \frac{1}{2} v^2 + \frac{\gamma}{\gamma - 1} \frac{P}{\rho} + \Phi\,,
\end{equation}
where $\Phi$ is the gravitational potential, $P$ the gas pressure, $\rho$ the gas density, and $v$ the gas velocity, and mark it in Fig.~\ref{fig:densityandvelocity} with a dashed red line.
The bound region extend with a long tail close to the disc midplane, but from the cut-off radius outwards the disc wind streamlines start to curve downwards, and not radially away from the star as in the extended disc case. As a result, the surface mass-loss profiles are truncated very close to the cut-off radius, as shown in Fig.~\ref{fig:mass_loss_rate_cumulative}.
Some material is recirculated outside the cut-off radius, as shown by the velocity field in Fig.~\ref{fig:densityandvelocity}. However, this effect is marginal at best. The amount of material recirculated can be estimated from the tail of the cumulative mass loss rate. Its effect might be more pronounced for cut-off radii smaller than 10 au where the cumulative mass-loss rate scales as $R^3$, but even in this case it would be on the order of $10^{-10}\, M_\odot\,\mathrm{yr}^{-1}$, enhanching only slightly the viscous expansion of the disc. If one takes into account environment effects, like external photoevaporation, this loosely bound material would be nevertheless removed from the system.

We computed the cumulative mass-loss rate from our hydrodynamical models, as shown in Fig.~\ref{fig:mass_loss_rate_cumulative}. The four cases exhibit very similar radial behaviours, with $\dot{M}_w(R)$ rising steeply in the inner disc and flattening towards large radii. This outcome is expected, since the incident radiation field at smaller radii is not directly affected by the details of the density distribution further out.
The main effect of reducing the cut-off radius is to truncate the radial extent of the mass-loss distribution (equation~\ref{eq:surf2}), thereby lowering the total integrated mass-loss rate.
In Fig.~\ref{fig:mass_loss_rate_cumulative} we show also that the complex fit of equation~\ref{eq:surf2} can be actually broken in two linear regimes, in order to find a simpler (and physically motivated) way to describe the wind mass-loss, where for radii comparable with the gravitational radius it grows as $\dot{M}_w \propto R^3$, while for larger radii like $\dot{M}_w \propto R^{0.5}$. This shows how important it is to consider the cut-off radius truncated mass-loss rate for compact discs, where the bulk of the disc fits inside the fast growth regime of the wind mass-loss rate.
Our heating prescription based on the ionization parameter assumes that the main source of heating is direct irradiation. This assumption should hold in most of the cases, as the diffuse material close to the disc outer edge has a very low density and the resulting diffuse EUV field is negligible compared to environmental effects. Thus we don't expect an increase in the mass-loss rate outside the cut-off radius from internal photoevaporation, but we explore the effect of external photoevaporation in the population synthesis models in the next section.

We can now provide a prescription for the long-term evolution of compact discs. The normalization factor to be used is given by subtituting in equation~\ref{eq:surf2} the cut-off radius $r_1$
\begin{equation}\label{eq:mdotnorm}
    \dot{M}_{w, \mathrm{norm}}(R) = \dot{M}_{w, \mathrm{norm}}(r_1)\,.
\end{equation}
The consistency of the surface mass loss rate distribution however is reassuring, as it demonstrates that existing mass-loss prescriptions remain valid for compact discs, provided they are appropriately scaled to the total wind rate according to the relation above. In the following section, we normalize the mass-loss rates to the value of the cumulative mass-loss rate at the cut-off radius.

\subsection{Single disc evolution}

\begin{figure*}
    \includegraphics{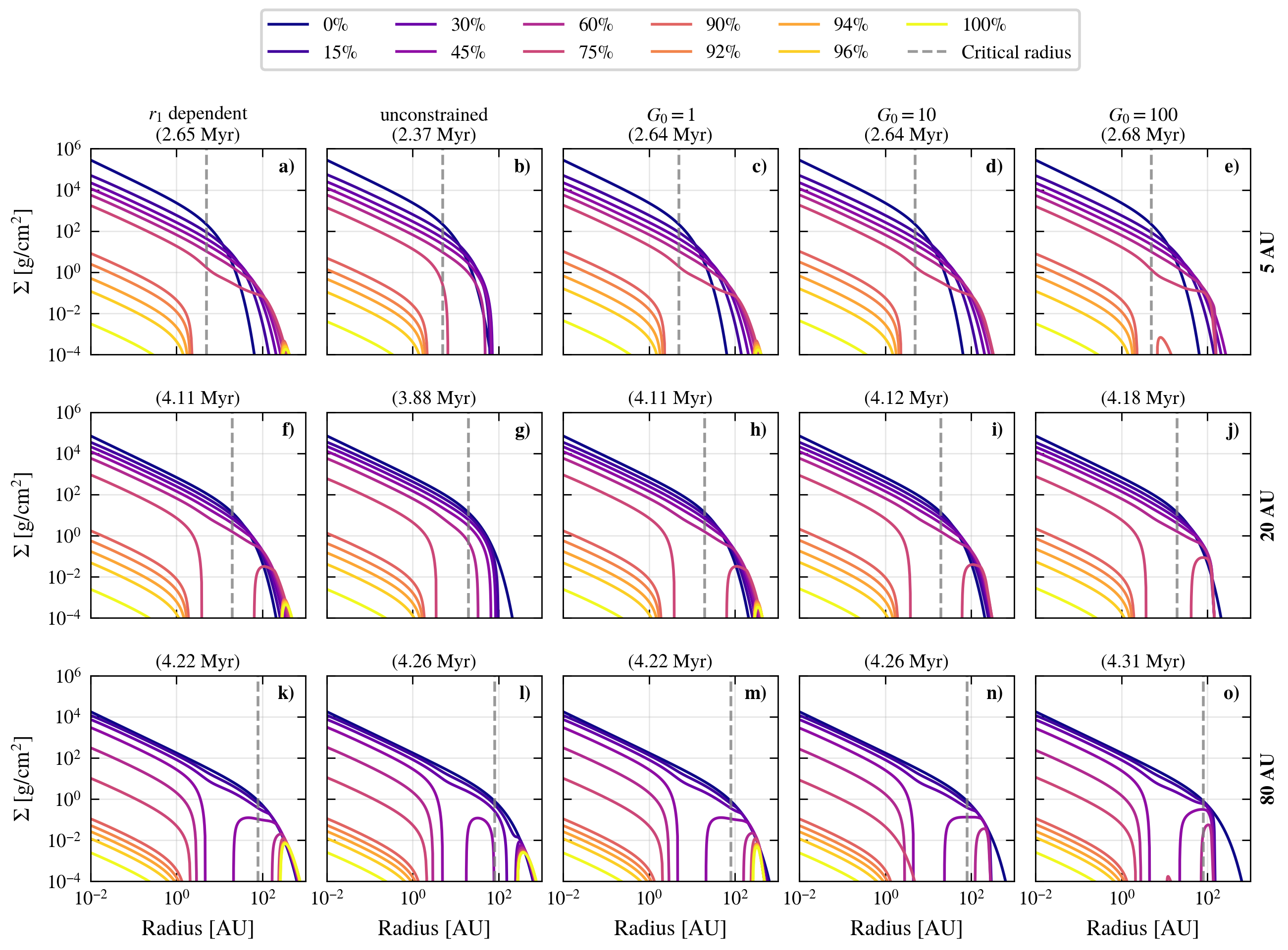}
    \caption{Surface density evolution for discs with a cut-off radius of 5 au (top row), 20 au (middle row) and 80 au (bottom row) for the internal photoevaporation limited to the cut-off radius (left column) and the unconstrained internal photoevaporation prescription (right column). The line colours indicate the different evolutionary stage as explained in the legend on top.}
    \script{fig4.py}
    \label{fig:photoevaporation_scenarios}
\end{figure*}

We investigated the disc long-term evolution using our 1-D viscous evolution code \textsc{spock} for discs with $r_1 = 5, 20,$ and $80$ au, $M_\mathrm{d} = 0.01\, M_\odot$, $\alpha=10^{-3}$ around a \SI{1}{M_\odot} star with $L_X = 2.29\cdot10^{30}$ erg/s, evolving the surface density according to equations~\ref{eq:visc}-\ref{eq:LxMstar} and \ref{eq:mdotnorm} with a factor $10$ decrease in the internal photoevaporation rate.
The cut-off radius is updated at each timestep.
In Figure~\ref{fig:photoevaporation_scenarios}, the first column shows the surface density evolution obtained adopting the cut-off dependent mass-loss prescription, while the second one displays the corresponding results using the standard, unscaled prescription.
The two approaches lead to strikingly different evolutionary outcomes.
For compact discs (with cut-off radius less than 20-30 au) with unconstrained internal photoevaporation, the wind mass-loss rate is stronger than the viscous spreading outside the cut-off radius.
As a result, these discs tend to disperse from the outside-in, gradually eroding their outer edges until the disc vanishes.
When the disc is not initially compact (bottom row) then the two prescriptions yield similar results. The disc dispers generally from the inside-out with a cavity being opened in the outer disc depending on the initial surface density profile.
In contrast, when the cut-off dependence is included, the compact discs evolve via inside-out dispersal, with gap opening driving the final dispersal.
This distinction is not only physically motivated, but also observationally significant. The inside-out dispersal is consistent with the colour-colour evolution of young stellar objects (YSOs) observed in large surveys \citep{2013MNRAS.428.3327K}, which strongly disfavour outside-in clearing. The improved treatment of compact discs therefore provides a natural explanation for the observational evidence, reconciling theory and data.
For increasing external photoevaporation fields (from column 3 to 5), the disc is prevented to spread - as in the case of the unconstrained internal photoevaporation - but the disc dispersal happens still in an inside-out fashion, for the parameters explored here.

\subsection{Population synthesis models}

\begin{figure}
    \begin{centering}
	\includegraphics{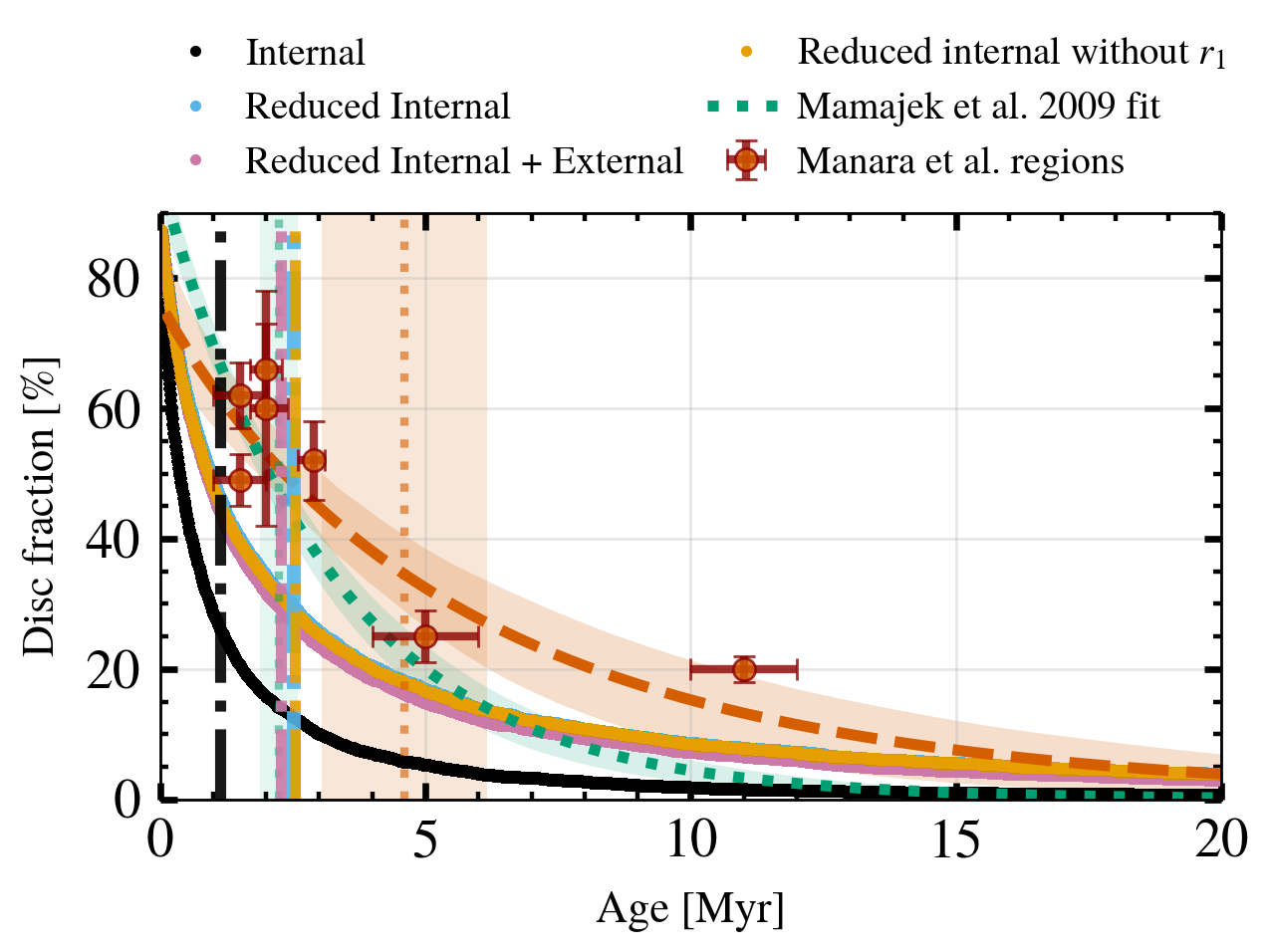}
	\caption{Disc fraction as a function of time for the different models explored. For comparison, the exponential fit from \citet{2009AIPC.1158....3M} is shown with a green dashed line and the the disc fractions of the star forming regions in \citet{2023ASPC..534..539M} is shown with orange dots. The median disc lifetime is overplotted with vertical dotted lines for each model and the observations.}
        \script{fig5_mamajek.py}
	\label{fig:disc_fraction_age}
    \end{centering}
\end{figure}

We performed a population synthesis of $N_{\rm disc} = 4096$ discs, with initial cut-off radii and masses drawn from the observationally derived distribution described in Section~\ref{sec:methods}.
The resulting population naturally reproduces the observed disc fraction as a function of cluster age \citep[e.g.][]{2009AIPC.1158....3M} for the reduced internal photoevaporation prescription without the need of tuning the parameters as shown in Figure~\ref{fig:disc_fraction_age}. The addition of external photoevaporation leads to a slightly faster decline of the disc fraction, but still compatible with the observations.
The resulting median disc lifetime is $2.70 \pm 0.13$ Myr, which is compatible with the observationally derived one ($2.24 \pm 0.35$ Myr) and considerably larger than that predicted without decreasing the internal photoevaporation rate ($1.13 \pm 0.05$ Myr).
The cut-off radius prescription has a limited effect on the overall disc lifetime, as the bulk of the disc mass is located at smaller radii where the mass-loss rates are similar in both models.
When considering only star forming regions within $200$ pc in the \citet{2023ASPC..534..539M} sample, the observed median disc lifetime is $4.61 \pm 1.55$ Myr, which is considerably longer than what predicted by current models but more in line with the improved prescription for compact discs.

\begin{figure*}
    \begin{centering}
        \includegraphics{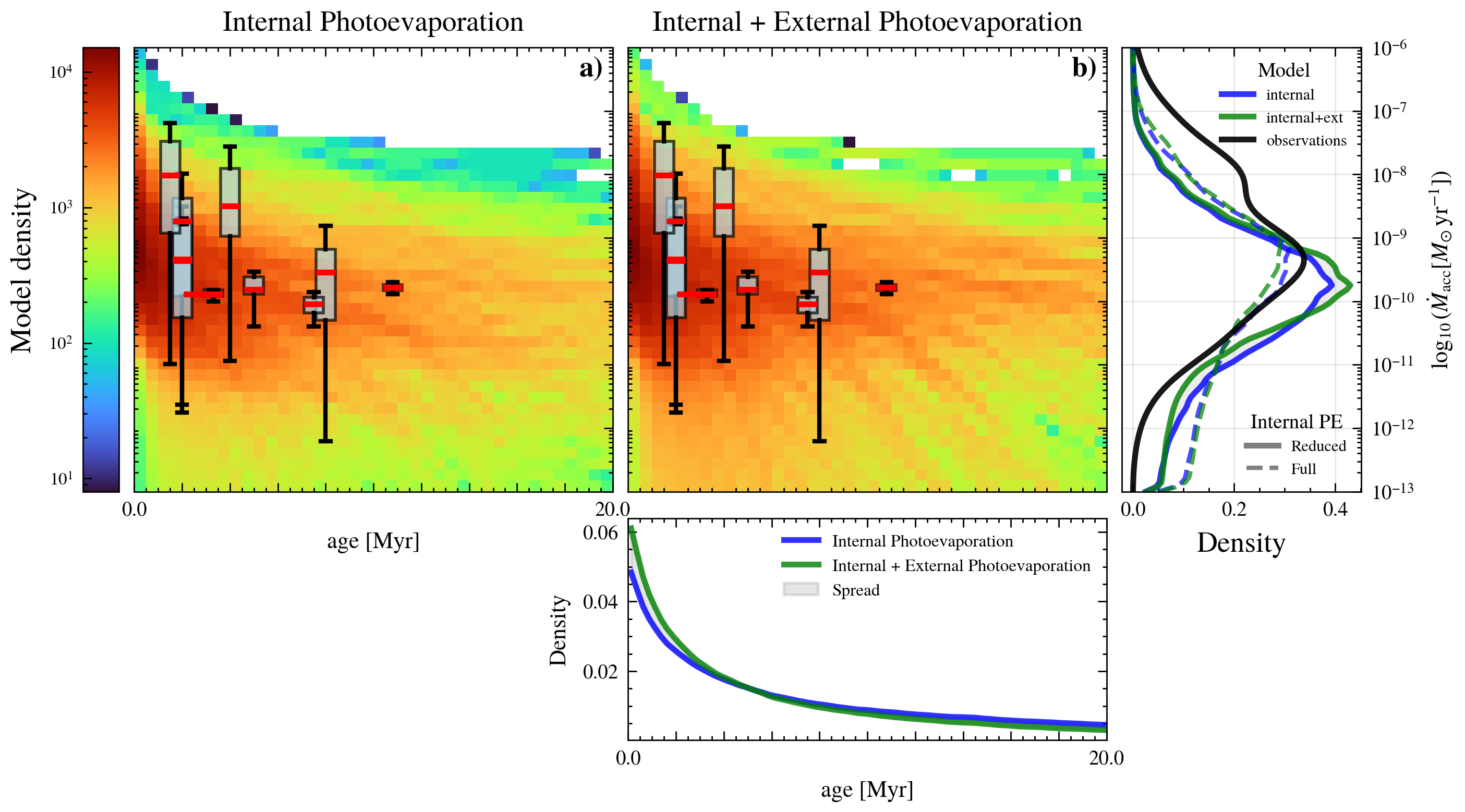}
        \caption{Mass accretion rate as a function of time for the old prescription (panel a) and the cut-off radius criteria (panel b). The corner plots show the density distribution in the accretion rates and disc age, and the difference as a gray shadow region. The disc observed properties \citep{2023ASPC..534..539M, 2022yCat..51630074T} are overplotted as box plots.}
        \label{fig:mass_accretion_age}
        \script{fig6.py}
    \end{centering}
\end{figure*}

We then compared the accretion rate distributions as a function of age predicted by our population synthesis models with observational data from several stellar clusters, including Taurus, Lupus, Upper Scorpius, Chameleon I, Chameleon II, $\rho$ Ophiuchi, Corona Australis, Orionis Cloud A, Orionis Cloud B, $\gamma$ Vel, $\sigma$ Orionis, OriOB1a, OriOB1b \citep{2023ASPC..534..539M, 2022yCat..51630074T}. 
We evaluated the predictions obtained using the cut-off dependent mass-loss prescription and including or not the effect of external photoevaporation.
As the population synthesis models with reduced internal photoevaporation (by a factor of 10) better reproduce the observed disc lifetimes, we focus the rest of our analysis on this set of models, but we keep a comparison with the standard prescription.
Both models correctly predict the observed general evolution of the mass-accretion rates, as shown in Fig.~\ref{fig:mass_accretion_age}.
The effect of external photoevaporation is to reduce slightly the disc lifetime and thus the population of low-accreting systems.
Compared to the observations, the models underpredict a population of high accreting systems which could be explained by an initial absorption of the high energetic stellar irradiation by a massive inner MHD wind \citep{2023ASPC..534..567P}.
When comparing the reduced internal photoevaporation prescription with the standard one marked with dashed lines in the right panel, there is a slight shift of the peak towards lower accretion rates, and a lower fraction of low accreting systems, which is more in line with the observations.

\begin{figure*}
	\centering
	\includegraphics{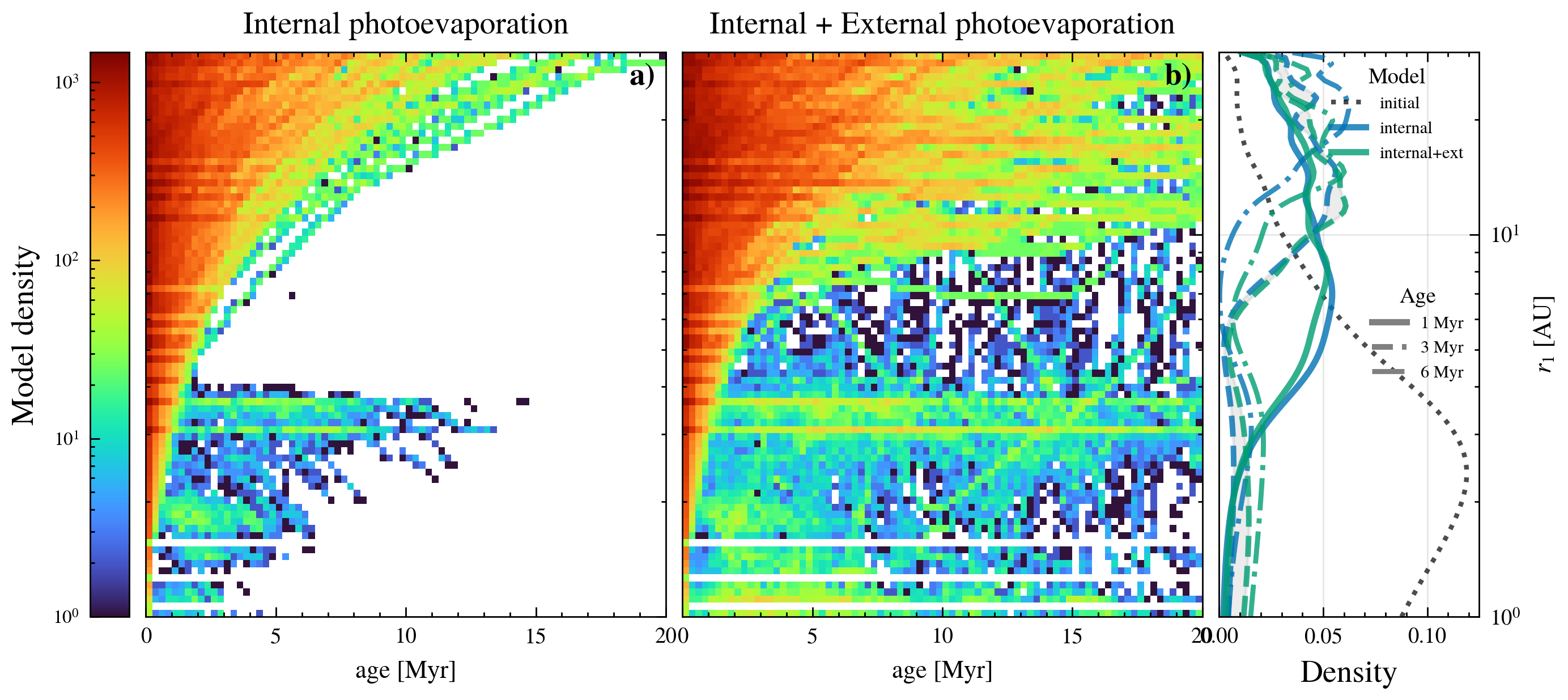}
	\caption{Cut-off radius as a function of time for internal (panel a) and internal+external photoevaporation (panel b) for compact discs ($r_1 < 30$ au). The right panel show the cut-off radius density distribution at different ages for the two models.}
	\label{fig:critical_radius_age}
    \script{fig7_compact.py}
\end{figure*}

Finally we focused on the effect of the external photoevaporation on the cut-off radius evolution for compact discs (with initial $r_1 < 30$ au), as shown in Fig.~\ref{fig:critical_radius_age}.
Viscous spreading leads to a rapid increase of the cut-off radius.
Internal photoevaporation, opening up a gap around the gravitational radius and dispersing the disc from the inside-out, hinders the disc radial expansion, as seen for a class of compact discs that keep a cut-off radius less than 5 au for more than 5 Myr.
The addition of external photoevaporation leads generally to a more gentle increase of the cut-off radius, and to a larger fraction of compact discs.
For larger external FUV fields, one can envision a scenario where the cut-off radius remains small for the entire disc lifetime, as viscous expansion drives the disc radii to an equilibrium radius where visous expansion and external photoevaporation cancel each other out.
For the parameter space explored here, the bulk of the disc population still experiences an increase of the cut-off radius over time, but the fraction of compact discs is larger than in the case without external photoevaporation, as shown in the right panel comparing at later ages the distribution of the population driven by internal photoevaporation only and by internal plus external photoevaporation.


\section{Conclusions}\label{sec:conclusions}

We have shown that compact discs, with a cut-off radius comparable with the gravitational radius for photoevaporation, follow the same local mass-loss profile as extended discs, but their integrated wind rates are reduced according to the previously found cumulative mass-loss rate distribution, given in equations~\ref{eq:mdotnorm}, \ref{eq:mdot},  \ref{eq:mdotmstar}, \ref{eq:mdotlx}, \ref{eq:lxsoft}.
Incorporating this scaling into viscous evolution models compact discs leads to longer disc lifetimes and inside-out dispersal, in agreement with observational constraints from colour-colour diagrams of young stellar populations.
Population synthesis taking into account the evolution of compact discs reproduces better the observed disc fractions and accretion rates as a function of time (Figures \ref{fig:disc_fraction_age}, \ref{fig:mass_accretion_age}).
Including the effect of external photoevaporation leads to a more gentle increase of the cut-off radius for compact discs, and a shift towards smaller cut-off radii at later ages (Figure~\ref{fig:critical_radius_age}).
The dispersal of compact discs cannot be captured by standard unscaled models, and radius-dependent photoevaporation prescriptions are essential to connect disc evolution theory with observations.


\section*{Acknowledgements}

The authors would like to thank the anonymous referee for the constructive comments and valuable suggestions, which helped improve the quality of this manuscript.
This work was supported by the Deutsche Forschungsgemeinschaft (DFG, German Research Foundation) Research Unit "Transition discs" - 325594231 and of the Excellence Cluster ORIGINS - EXC-2094 - 390783311.


\section*{Data availability}
This study uses the reproducibility framework ``showyourwork''~\citep{Luger2021}.
All code required to reproduce our results, figures, and this article itself is available at \url{https://github.com/GiovanniPicogna/disc-photoevaporation-disk-radii}.
The code to reproduce a figure can be accessed via the icon link next to the respective figure caption.
Data sets associated with this work are available at doi:10.5281/zenodo.20071852


\bibliographystyle{mnras}
\bibliography{biblio} 

\appendix

\bsp	
\label{lastpage}
\end{document}